# Effect of the Intrinsic Width on the Piezoelectric Force Microscopy of a Single Ferroelectric Domain Wall

Anna N. Morozovska [*]

Institute of Semiconductor Physics, National Academy of Science of Ukraine,

41, pr. Nauki, 03028 Kiev, Ukraine

Eugene A. Eliseev

Institute for Problems of Materials Science, National Academy of Science of Ukraine,

3, Krjijanovskogo, 03142 Kiev, Ukraine

George S. Svechnikov

Institute of Semiconductor Physics, National Academy of Science of Ukraine,

41, pr. Nauki, 03028 Kiev, Ukraine

Venkatraman Gopalan

Department of Materials Science and Engineering, Pennsylvania State University, University

Park, PA16802

Sergei V. Kalinin[†]

Materials Science and Technology Division and Center for Nanophase Materials Science,

Oak Ridge National Laboratory, Oak Ridge, TN 37831

---

[*] Corresponding author, morozo@i.com.ua

[†] Corresponding author, sergei2@ornl.gov

**Abstract**

Intrinsic domain wall width is a fundamental parameter that reflects bulk ferroelectric properties and governs the performance of ferroelectric memory devices. We present closed-form analytical expressions for vertical and lateral piezoelectric force microscopy (PFM) profiles for the conical and disc models of the tip, beyond point charge and sphere approximations. The analysis takes into account the finite intrinsic width of the domain wall, and dielectric anisotropy of the material. These analytical expressions provide insight into the mechanisms of PFM image formation and can be used for quantitative analysis of the PFM domain wall profiles. PFM profile of a realistic domain wall is shown to be the convolution of its intrinsic profile and resolution function of PFM.

**I. Introduction**

Piezoelectric Force Microscopy has become the technique of choice for nanoscale imaging and characterization (see e.g. Refs. 1, 2, 3), allowing for recent advances in PFM results interpretation.[4] Now PFM methods are widely used for manipulation and tailoring of ferroelectric domains structure (see e.g. Refs. 5, 6, 7), study of ferroelectric domain growth dynamics,[8] visualization and local characterization of capacitors structures,[9] local polarization switching,[10] polycrystalline and relaxor ferroelectrics,[11] and size effects in ultrathin ferroelectric films.[12] In order to interpret the experimental results of the PFM image of a ferroelectric domain wall with "sharp" tips (with high spatial resolution), one has to take into account the natural distribution of material properties such as the piezoelectric, dielectric and elastic coefficients across the domain wall. For most of bulk and thin films ferroelectrics, the intrinsic wall width ranges from one to several lattice constants, as it was recently found by Atomic Force Microscopy[13] and advanced Transmission Electron Microscopy.[14] Similar length-scale has been found in polarization distribution over the ferroelectric film thickness.[15] However, it was found recently that domain walls broaden on the surface compared to the bulk of a ferroelectric crystal.[16, 17] Furthermore, the presence of stresses or charge or dipolar defects can result in elastic, dielectrics, and electromechanical property gradients on the length-scales from 10 to 1000 nanometers. It is therefore essential to quantitatively understand the influence of the intrinsic properties distribution across a wall on the observable PFM profile of the wall measured with sharp tips.

Recent results[18, 19] of domain wall PFM imaging as well as the finite element modeling suggest that the intrinsic domain wall width has an effect on the measured PFM profile. Here we derive analytical expressions for vertical and lateral PFM profiles of finite-width 180° domain walls taking into account the contact and the conical parts of the probe, as well as dielectric anisotropy of the material.

The paper is organized as follows. Basic principles of PFM response calculation and electrostatic field structure of the probe are discussed in Section II. The relationship between the PFM profile of sharp domain walls and material properties is analyzed in Section III for conic, contact and effective point charge tip models. The influence of intrinsic width on domain wall PFM profile is considered in Section IV. Obtained results are discussed in Section V.

### II. Basic equations

In the case of the strain piezoelectric coefficient $d_{klj}$ dependent only on lateral coordinates (system may be considered uniform in $z$-direction), the surface displacement vector $u_i(\mathbf{y})$ (measured PFM piezoresponse) is given by the convolution of an *ideal image* $d_{klj}(\mathbf{y}-\mathbf{x})$ with the resolution function components $W_{ijkl}(\mathbf{x})$ (see Ref. [20]). Since in many cases, the inhomogeneous distribution of piezoelectric coefficients are similar, e.g. for ferroelectrics they are determined by polarization distribution, hereinafter we introduce the inhomogeneous part of piezoelectric coefficients as $\beta(\mathbf{y}-\mathbf{x})$, i.e. assume that $d_{klj}(\mathbf{y}-\mathbf{x}) \equiv d_{klj} \cdot \beta(\mathbf{y}-\mathbf{x})$. Here $d_{klj}$ are the piezoelectric tensor components of homogeneous media; absolute value of function $\beta(\mathbf{y}-\mathbf{x})$ is smaller than unity. Dielectric permittivity and elastic modules are regarded constant within the sample. In this approximation, components of the surface displacement below the tip can be written as follows:[20]

$$u_i(\mathbf{y}) = \int_{-\infty}^{\infty} dx_1 \int_{-\infty}^{\infty} dx_2 \, W_{ijkl}(-x_1,-x_2) d_{lkj} \beta(y_1-x_1, y_2-x_2), \qquad (1)$$

where the resolution function is introduced as:

$$W_{ijkl}(x_1,x_2) = c_{kjmn}\int_0^\infty dx_3 \frac{\partial G_{im}(-x_1,-x_2,x_3)}{\partial x_n} E_l(x_1,x_2,x_3) \quad (2)$$

and $E_l$ is the component of the external electric field produced by the probe.

Here we calculate surface displacement below the tip located near the plain domain wall (see Fig. 1).

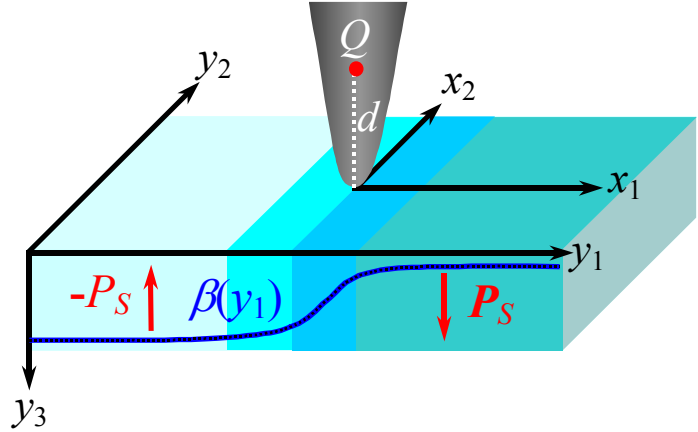

**FIG. 1**. (Color online). Schematics of PFM measurement across 180°-domain wall between the domains with opposite polarization $\pm P_S$.

For the infinitely thin isolated domain wall Eq. (1) can be rewritten as:

$$u_i^{step}(y_1) = \int_{-\infty}^{\infty} dx_1 \int_{-\infty}^{\infty} dx_2\, W_{ijkl}(-x_1,-x_2)d_{lkj}\,\text{sign}(y_1 - x_1) \quad (3)$$

Using the Delta function properties in Eq. (3), the displacement components for the arbitrary one-dimensional distribution $d_{lkj}\beta(y_1)$, i.e. measured wall profile, is

$$u_i(y_1) = \frac{1}{2}\int_{-\infty}^{\infty} u_i^{step}(y_1 - x)\frac{\partial \beta(x)}{\partial x}dx . \quad (4)$$

Hereinafter $\beta(y_1)$ is the "intrinsic" domain wall profile (see Appendix A for details) due to symmetry breaking, surface, or defect- or strain effects on piezoelectric properties. Eq. (4) is the main result, that allows calculation of broadened wall profile, if the profile $u_i^{step}$ of infinitely thin domain wall is available.

Using decoupling approximation[21,22] and resolution function theory,[20] piezoelectric response of isolated 180°-domain wall in the inhomogeneous electric field of the probe tip is:

$$d_{33}^{eff} = \frac{u_3}{V} = g_{313} d_{31} + g_{333} d_{33} + g_{351} d_{15},$$
$$d_{34}^{eff} = u_2 = 0, \qquad (5)$$
$$d_{35}^{eff} = \frac{u_1}{V} = g_{113} d_{31} + g_{133} d_{33} + g_{151} d_{15}.$$

Here $V$ is electric bias applied to the probe tip and Voigt matrix notations are used for piezoelectric tensor components. The explicit form of the tensorial functions $g_{ijk}$ (and hence that of the PFM image) depends on the tip coordinates $\{y_1, y_2\}$, domain wall intrinsic structure and the electric field distribution of the probe. Piezoelectric response in homogeneous electric field (flat capacitor geometry) is considered elsewhere.[23]

The electrostatic field produced by the tip includes the contributions from the conical part of the probe as well as tip-surface contact area. The conical part can be approximated by a line charge (see e.g. Refs. 24, 25, 26), and the contact area could be modeled by a disk touching the sample surface, as proposed in Ref. [18]. Finally, spherical part of the probe can be approximated by a point charge. Using superposition principle, we represent the probe electrostatic potential, $\varphi$, as the sum of effective line charge potential, $\varphi_L$, point charge potential, $\varphi_q$, and disk potential, $\varphi_D$:

$$\varphi(\rho, z) = \varphi_L(\rho, z) + \varphi_q(\rho, z) + \varphi_D(\rho, z), \qquad (6)$$

where cylindrical coordinates $\rho = \sqrt{x_1^2 + x_2^2}$ and $z \equiv x_3$ are introduced. Normalization in Eq. (6) requires $\varphi_L(0,0) + \varphi_D(0,0) + \varphi_q(0,0) = V$, corresponding to ideal electrical contact between the tip and the surface. Under the condition $\varepsilon_e \ll \varepsilon_{11,33}$ typically valid for the majority of ferroelectrics with $\varepsilon_{11,33} \geq 100$ and $\varepsilon_e < 10$, we obtained expressions for potential structure (see Appendixes B and C):

$$\varphi_L(\rho,z) = \frac{V}{\ln(\mathrm{ctg}^2\theta/2)} \begin{cases} \frac{2\varepsilon_e}{\varepsilon_e+\kappa}\ln\left(\frac{L+\Delta L+z+\sqrt{(L+\Delta L+z)^2+\rho^2}}{\Delta L+z+\sqrt{(\Delta L+z)^2+\rho^2}}\right), & z>0 \\ \ln\left(\frac{L+\Delta L+z+\sqrt{(L+\Delta L+z)^2+\rho^2}}{\Delta L+z+\sqrt{(\Delta L+z)^2+\rho^2}}\right) + \\ +\frac{\varepsilon_e-\kappa}{\varepsilon_e+\kappa}\ln\left(\frac{L+\Delta L-z+\sqrt{(L+\Delta L-z)^2+\rho^2}}{\Delta L-z+\sqrt{(\Delta L-z)^2+\rho^2}}\right), & z<0 \end{cases} \quad (7a)$$

$$\varphi_q(\rho,z) = \frac{q^*}{4\pi\varepsilon_0\varepsilon_e} \begin{cases} \frac{2\varepsilon_e}{\varepsilon_e+\kappa}\frac{1}{\sqrt{(\Delta L+z)^2+\rho^2}}, & z>0 \\ \frac{1}{\sqrt{(\Delta L+z)^2+\rho^2}} + \frac{\varepsilon_e-\kappa}{\varepsilon_e+\kappa}\frac{1}{\sqrt{(\Delta L-z)^2+\rho^2}}, & z<0 \end{cases} \quad (7b)$$

$$\varphi_D(\rho,z) = \frac{2V}{\pi}\arcsin\left(\frac{2R_0}{\sqrt{(\rho-R_0)^2+z^2}+\sqrt{(\rho+R_0)^2+z^2}}\right) \times$$
$$\times \left(1 - \frac{2\varepsilon_e}{\varepsilon_e+\kappa}\left(\frac{\ln(1+L/\Delta L)}{\ln(\mathrm{ctg}^2\theta/2)} + \frac{q^*}{4\pi\varepsilon_0\varepsilon_e V\Delta L}\right)\right) \quad (7c)$$

Hereinafter $\kappa = \sqrt{\varepsilon_{33}\varepsilon_{11}}$ is the effective dielectric constant and $\gamma = \sqrt{\varepsilon_{33}/\varepsilon_{11}}$ is the dielectric anisotropy factor, $\varepsilon_e$ is the ambient dielectric constant. The conical part potential Eq. (7a) is modeled by the linear charge of length $L$ with a constant charge density $\lambda_L = 4\pi\varepsilon_0\varepsilon_e V/\ln(\mathrm{ctg}^2\theta/2)$,[24] where $\theta$ is the cone apex angle. Additional point charge potential (7b) is chosen to reproduce the conductive tip surface as closely as possible by the isopotential surface $\varphi(\rho,z) = V$. Our numerical calculations show that the charge $q^*$ is located at the end of the line, at that the distance $\Delta L$ from the surface is approximately equal to the disk radius $R_0$ and $q^* = 4\pi\varepsilon_0\varepsilon_e V\Delta L$ for a typical range of cone angles $\theta$. Contact area potential Eq. (7c) is modeled by the disk of radius $R_0$, with surface charge density $\sigma_d \sim VR_0/\sqrt{R_0^2-\rho^2}$, where $\rho$ is the radial coordinate (see Fig. 2a). The vertical scale in Fig. 2a is compressed, since $L \gg R_0$ for a real probe apex shape. It is clear from the Fig. 2b, that isopotential surface $\varphi(\rho,z) = V$ reproduces the conductive tip shape in the vicinity of surface.

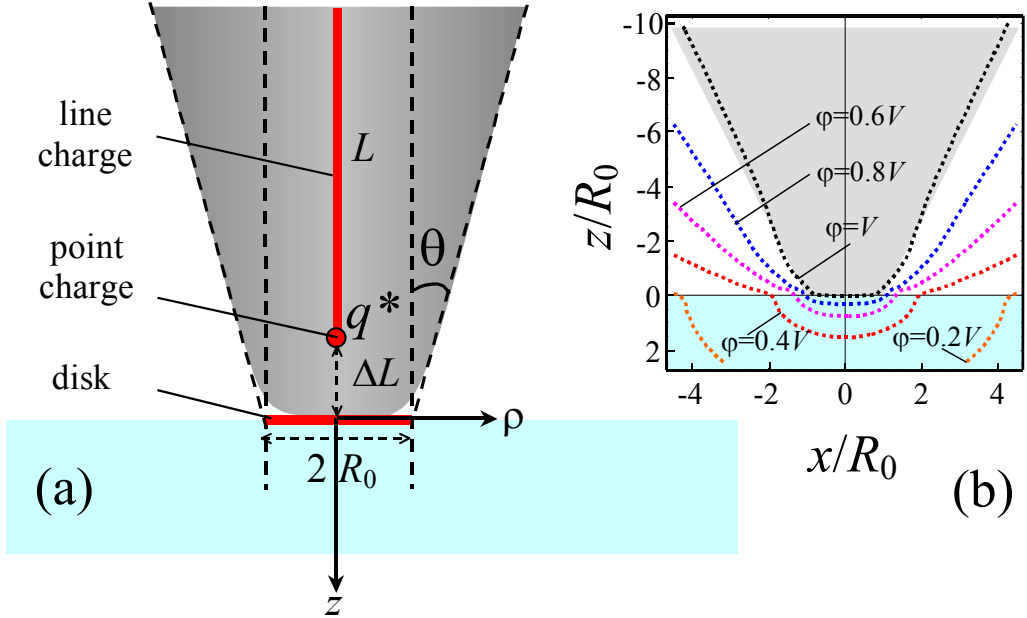

**FIG. 2**. (Color online). (a) Tip-surface contact. (b) Corresponding isopotential lines near the sample surface at $\Delta L = R_0$, $\theta = \pi/9$ and $q^* = 4\pi\varepsilon_0\varepsilon_e V \Delta L$.

It is important for further consideration that under the conditions $\Delta L = R_0$, $\rho \ll R_0$ and $0 \leq z$ the ratio of the conical terms (line + point charge) to the disk one can be estimated as

$$\frac{\varphi_q + \varphi_L}{\varphi_D} \leq \frac{2\varepsilon_e}{\varepsilon_e + \kappa} \frac{\ln(1 + L/R_0)}{\ln(\mathrm{ctg}^2 \theta/2)} \left(1 - \frac{2\varepsilon_e}{\varepsilon_e + \kappa} \frac{\ln(1 + L/R_0)}{\ln(\mathrm{ctg}^2 \theta/2)}\right)^{-1} \quad (7d)$$

So that it is small enough inside the sample under the typical condition $\varepsilon_e \sim 1$, $\kappa \geq 100$ and $L/R_0 \leq 10^3$, $\theta < \pi/4$, because $\kappa \gg \varepsilon_e$ and $L/R_0 \ll \exp((\varepsilon_e + \kappa)/2\varepsilon_e)$. At the same time, the condition $\kappa \gg \varepsilon_e$ is necessary for the validity of Eqs. (7). Numerical simulations prove that the relative contribution of conic potential $(\varphi_q + \varphi_L)$ at $z \leq 0$ is not more than 5% of the disk potential $\varphi_D$ (in agreement with the conclusion of Ref. [27]). However below we demonstrate that the conic part may affect on the piezoresponse saturation rate away from the domain wall.

It should be noted that the PFM profiles of domain walls for other tip models, like sphere-plane[27], can be considered on the basis of the results above by summation or integration of the point charges, representing the tip. The PFM profiles for more sophisticated domain structures with ideal walls, like periodic structures or cylindrical domain, were considered earlier for a spherical model of the tip.[20] Effective point charge approach is evolved in Ref. [28], neglecting the conic part contribution. Effective point charge approach uses only the point charge potential given by Eq. (7b) with $\Delta L \to h$, where effective distance $h$ is proportional to tip radius and depends on the tip model. In particular, the effective charge-surface separation is $h = 2R_0/\pi$ for a disk model or $h \approx 2\varepsilon_e R_0 \ln((\varepsilon_e + \kappa)/2\varepsilon_e)/(\kappa - \varepsilon_e)$ for the sphere-plane one.

### III. The influence of extrinsic factors on the width of domain wall PFM profile

The role of extrinsic factors on the broadening of PFM image of a domain wall can be demonstrated by the PFM profile width of an infinitely sharp flat domain wall. In fact the width of $d_{33}^{eff}(y)$ profile is exactly extrinsic (i.e. measured) domain wall width as determined by the PFM object transfer function (OTF) finite halfwidth,[20] that is defined by the tip size and geometry.

Using Eqs. (3, 5-7), displacement $d_{33}^{eff}(y)$ at the distance $y$ from an infinitely sharp domain wall were found as follows:

$$d_{3i}^{eff}(y) = \frac{u_i^{step}(y)}{V} = \left( \frac{\varphi_q(0,0)}{V} g_{ijk}^q(y) + \frac{\varphi_D(0,0)}{V} g_{ijk}^D(y) + \frac{\varphi_L(0,0)}{V} g_{ijk}^L(y) \right) d_{kj}. \tag{8}$$

Functions $g_{ijk}^q(y, \Delta L)$, $g_{ijk}^D(y, R_0)$ and $g_{ijk}^L(y, \Delta L, L)$ could be presented via universal function $g_{ijk}^S(y, h_S)$ at $s = q, D, L$ (see Appendixes B and C). For vertical piezoresponse, the nonzero displacement components are:

$$g_{313}^S(y, h_S) = (1+\nu)f_{313} \cdot p_3^S\left(\frac{y}{C_{313}h_S}\right) - f_{333} \cdot p_3^S\left(\frac{y}{C_{333}h_S}\right), \tag{9a}$$

$$g_{333}^S(y, h_S) = f_{333} \cdot p_3^S\left(\frac{y}{C_{333}h_S}\right), \qquad g_{351}^S(y, h_S) = f_{351} \cdot p_3^S\left(\frac{y}{C_{351}h_S}\right), \tag{9b}$$

For the lateral piezoresponse, the nonzero components are

$$g_{113}^S(y, h_S) = (1+\nu)f_{113} \cdot p_1^S\left(\frac{y}{C_{113}h_S}\right) - f_{133} \cdot p_1^S\left(\frac{y}{C_{133}h_S}\right), \tag{9c}$$

$$g_{133}^S(y,h_S) = f_{133} \cdot p_1^S\left(\frac{y}{C_{133}h_S}\right), \qquad g_{151}^S(y,h_S) = f_{151} \cdot p_1^S\left(\frac{y}{C_{151}h_S}\right), \qquad (9d)$$

Distances $h_q = h_L \approx \Delta L$ and $d_D = R_0$. Functions $p_3^S(\xi)$ and $p_1^S(\xi)$ represent the PFM image broadening of the ideal sharp domain wall. Their approximate expressions are:

$$p_3^q(\xi) = \frac{\xi}{|\xi|+1}, \qquad p_1^q(\xi) = \frac{1}{|\xi|+1}, \qquad (10a)$$

$$p_3^D(\xi) = \frac{2}{\pi}\arctan(\xi), \qquad p_1^D(\xi) = \frac{2}{\pi}\arctan\left(\frac{1}{|\xi|}\right), \qquad (10b)$$

$$p_3^L(\xi) = \frac{\text{sign}(\xi)}{\ln(1+\varsigma)}\ln\left(\frac{(|\xi|+1)(1+\varsigma)}{|\xi|+1+\varsigma}\right), \qquad p_1^L(\xi) = \frac{\ln(|\xi|+1+\varsigma) - \ln(|\xi|+1)}{\ln(1+\varsigma)} \qquad (10c)$$

Here $\nu$ is the Poisson ratio and $\varsigma = L/\Delta L$. Constants $C_{ijk}$ and $f_{ijk}$ depend on the dielectric anisotropy factor $\gamma$ only and are listed in Appendix D. Constants $f_{3jk}$ determine the value of vertical response $d_{33}^{eff}(y)$ far from wall at $|y| \to \infty$. Constants $f_{1jk}$ determine the value of lateral response $d_{35}^{eff}(y)$ at $|y| \to 0$. Constants $C_{ijk}$ determine the effective width of PFM image of domain wall. It is obvious that when one of the terms $f_{1jk}d_{kj}$ or $f_{3jk}d_{kj}$ dominates the others, the halfwidth at half maximum (saturation level) for the lateral or vertical PFM profile will be $C_{1jk}h_S$ or $C_{3jk}h_S$ respectively. Material parameters, constants $f_{ijk}$ and $C_{ijk}$ for different ferroelectrics are listed in Tables I-III.

**Table I.** Ferroelectric material parameters

|  | $\varepsilon_{11}$ | $\varepsilon_{33}$ | $\gamma$ | $d_{15}$ (pm/V) | $d_{31}$ (pm/V) | $d_{33}$ (pm/V) |
|---|---|---|---|---|---|---|
| LiNbO$_3$ | 85 | 29 | 0.58 | 68 | -1 | 6 |
| LiTaO$_3$ | 54 | 44 | 0.90 | 26 | -2 | 8 |
| BaTiO$_3$ | 2920 | 168 | 0.24 | 392 | -34.5 | 85.6 |
| PbTiO$_3$ | 140 | 105 | 0.87 | 61 | -25 | 117 |

**Table II.** Vertical piezoresponse constants

|         | $f_{313}$ | $f_{333}$ | $f_{351}$ | $C_{313}$ | $C_{333}$ | $C_{351}$ |
|---------|-----------|-----------|-----------|-----------|-----------|-----------|
| LiNbO$_3$ | -1.26     | -0.86     | -0.14     | 0.24      | 0.21      | 0.68      |
| LiTaO$_3$ | -1.05     | -0.775    | -0.225    | 0.25      | 0.24      | 0.74      |
| BaTiO$_3$ | -1.61     | -0.96     | -0.04     | 0.19      | 0.12      | 0.51      |
| PbTiO$_3$ | -1.07     | -0.78     | -0.215    | 0.25      | 0.24      | 0.735     |

**Table III.** Lateral piezoresponse constants

|         | $f_{113}$ | $f_{133}$ | $f_{151}$ | $C_{113}$ | $C_{133}$ | $C_{151}$ |
|---------|-----------|-----------|-----------|-----------|-----------|-----------|
| LiNbO$_3$ | 1.46      | 0.47      | 0.17      | 0.27      | 0.31      | 0.92      |
| LiTaO$_3$ | 1.08      | 0.39      | 0.24      | 0.26      | 0.33      | 0.95      |
| BaTiO$_3$ | 2.40      | 0.57      | 0.06      | 0.27      | 0.22      | 0.79      |
| PbTiO$_3$ | 1.11      | 0.40      | 0.235     | 0.26      | 0.33      | 0.95      |

One can see from Eqs. (9-10) that the vertical response is zero at the center of the wall (at $y=0$) and saturates far from wall (at $|y|\to\infty$), while the lateral PFM response is maximal in the center of the wall and tends to zero far from wall.

Vertical PFM response near an infinitely thin domain wall in PbTiO$_3$ as a function of the distance from the wall is shown in Figs. 3.

Since $R_0 \approx \Delta L$, for the ratios $L/R_0 << \exp((\varepsilon_e + \kappa)/2\varepsilon_e)$, the relative contribution of conic part to vertical piezoresponse $d_{33}^{eff}(y)$ is generally no more than 5%, while the disk contribution strongly dominates (compare solid, dashed and dash-dotted curves in Fig.3 and see Eq.(7d)). The same is true for a lateral PFM response $d_{35}^{eff}(y)$ (not shown). However, the conic part does affect the saturation rate of the piezoresponse from the domain wall due to the long-range character of the produced electric fields.

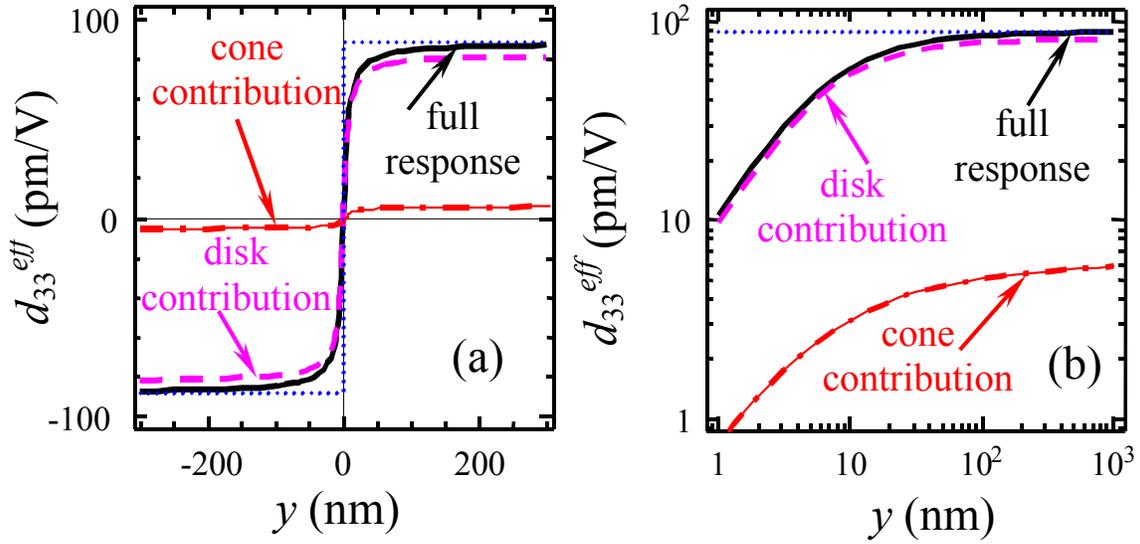

**FIG. 3.** (Color online) Vertical PFM response near the infinitely thin domain wall in PbTiO$_3$ as a function of the distance from the wall in linear scale (a) and log-log scale (b). Solid, dashed and dash-dotted curves are the full response, disk and cone contributions correspondingly. PbTiO$_3$ material parameters $\nu = 0.35$, $\kappa = 121$, $\gamma = 0.87$, $d_{33}^S = 117$, $d_{15}^S = 61$, $d_{31}^S = -25$ pm/V, ambient permittivity $\varepsilon_e = 1$, tip characteristics $R_0 = \Delta L = 20$ nm, $L = 2\,\mu$m, $\theta = \pi/9$.

Since the contribution of conic part is negligible in the vicinity of a domain wall $|y| < 2R_0$ (see e.g. Fig. 3a), the effective point charge approximation for a disk tip with $h = 2R_0/\pi$ and $s = q$ is valid in the region $|y| < 2R_0$ with satisfactory accuracy under the aforementioned condition $L/R_0 \ll \exp((\varepsilon_e + \kappa)/2\varepsilon_e)$ (see comments to Eq.(7d)). Numerical simulations prove that the same inequality for conic part length $L$ should be valid for effective point charge approximation of sphere-plane tip with curvature $R_0$.

Relative contribution of conical part becomes comparable with the disc one under the condition $L/R_0 \geq \exp((\varepsilon_e + \kappa)/2\varepsilon_e)$. Figs.4. illustrate the region of tip geometric parameters where the disk part (i.e. contact) contribution into the piezoresponse dominates over the conic one. Obtained curves are well described by the dependence $\theta = \pi - 2\,\mathrm{arctg}\left((1 + L/R_0)^{\varepsilon_e \Psi/(\varepsilon_e + \kappa)}\right)$ obtained directly from Eq.(7d), where the constant $\Psi \approx 6-10$. Note that narrow "tapered" probes heavily favor local contribution to the signal, while probes with large opening angles favor non-local contribution.

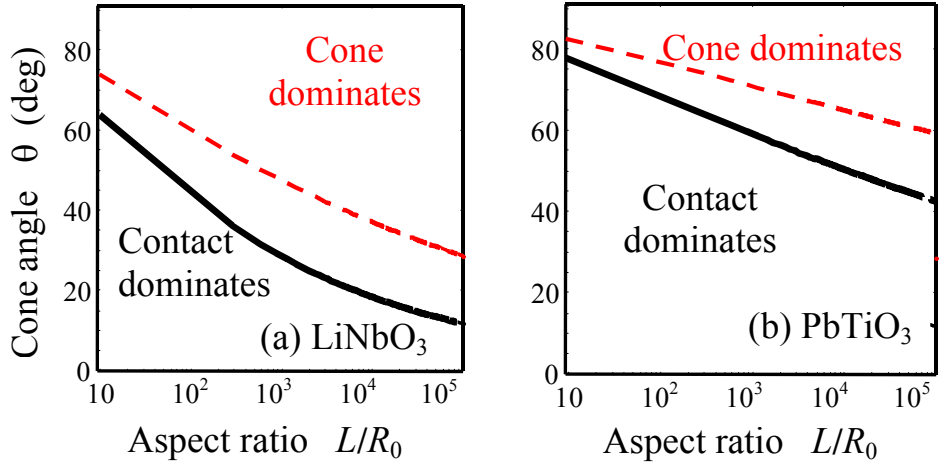

**FIG. 4.** (Color online) Region of tip geometric parameters contribution into the piezoresponse at $\varepsilon_e = 1$, $R_0 \approx \Delta L$, (a) LiNbO$_3$ and (b) PbTiO$_3$ material parameters. Solid line corresponds to the contact contribution of 95%, while the region above dotted line corresponds to the cone contribution more than 50%.

## IV. The influence of intrinsic width on domain wall PFM profile

The role of intrinsic factors on the broadening of the PFM image of a domain wall will now be considered for a domain wall with nonzero intrinsic (or natural) width. In order to consider the influence of intrinsic domain wall width analytically, we approximate $d_{kj}$ distribution as the oblique step $\beta(x,b) = (|x+b|-|x-b|)/2b$, where $2b$ is the intrinsic width of the wall. Using Eqs.(4), (5)-(7), piezoresponse $d_{3i}^{eff}(y,b)$ at the distance $y$ from the domain wall were found as follows:

$$d_{3i}^{eff}(y,b) = \frac{u_i(y,b)}{V} = \left( \frac{\varphi_q(0,0)}{V} g_{ijk}^q(y,b) + \frac{\varphi_D(0,0)}{V} g_{ijk}^D(y,b) + \frac{\varphi_L(0,0)}{V} g_{ijk}^L(y,b) \right) d_{kj} \quad (11)$$

Functions $g_{ijk}^q(y,b,\Delta L)$, $g_{ijk}^D(y,b,R_0)$ and $g_{ijk}^L(y,b,\Delta L, L)$ could be presented via universal function $g_{ijk}^m(y,b,h_m)$ at $m=q,D,L$. For vertical piezoresponse, the nonzero components are:

$$g_{313}^m(y,b,h_m) = (1+\nu)f_{313}p_3^m\left(\frac{y}{C_{313}h_m},\frac{b}{C_{313}h_m}\right) - f_{333}p_3^m\left(\frac{y}{C_{333}h_m},\frac{b}{C_{333}h_m}\right), \quad (12a)$$

$$g_{333}^m(y,b,h_m) = f_{333}p_3^m\left(\frac{y}{C_{333}h_m},\frac{b}{C_{333}h_m}\right), \quad g_{351}^m(y,b,h_m) = f_{351}p_3^m\left(\frac{y}{C_{351}h_m},\frac{b}{C_{351}h_m}\right), \quad (12b)$$

For the lateral piezoresponse, the nonzero components are:

$$g_{113}^m(y,b,h_m) = (1+\nu)f_{113}p_1^m\left(\frac{y}{C_{113}h_m},\frac{b}{C_{113}h_m}\right) - f_{133}p_1^m\left(\frac{y}{C_{133}h_m},\frac{b}{C_{133}h_m}\right), \quad (12c)$$

$$g_{133}^m(y,b,h_m) = f_{133}p_1^m\left(\frac{y}{C_{133}h_m},\frac{b}{C_{133}h_m}\right), \quad g_{151}^m(y,b,h_m) = f_{151}p_1^m\left(\frac{y}{C_{151}h_m},\frac{b}{C_{151}h_m}\right). \quad (12d)$$

Distances $h_q = h_L \approx \Delta L$ and $h_D = R_0$. Functions $p_3^m(\xi,\omega)$ and $p_1^m(\xi,\omega)$ represent the PFM image broadening of an oblique step domain wall:

$$p_3^q(\xi,\omega) = \frac{|\xi+\omega|-|\xi-\omega|}{2\omega} + \frac{1}{2\omega}\ln\left(\frac{|\xi-\omega|+1}{|\xi+\omega|+1}\right), \quad (13a)$$

$$p_1^q(\xi,\omega) = \text{sign}(\xi+\omega)\frac{\ln(|\xi+\omega|+1)}{2\omega} - \text{sign}(\xi-\omega)\frac{\ln(|\xi-\omega|+1)}{2\omega}, \quad (13b)$$

$$p_3^D(\xi,\omega) = \frac{2}{\pi}\left(\begin{array}{l}\arctan(\xi+\omega)\dfrac{\xi+\omega}{2\omega} - \arctan(\xi-\omega)\dfrac{\xi-\omega}{2\omega} + \\ +\dfrac{1}{4\omega}\ln\left(\dfrac{(\xi-\omega)^2+1}{(\xi+\omega)^2+1}\right)\end{array}\right), \quad (13c)$$

$$p_1^D(\xi,\omega) = \frac{2}{\pi}\left(\begin{array}{l}\arctan\left(\dfrac{1}{|\xi+\omega|}\right)\dfrac{\xi+\omega}{2\omega} - \arctan\left(\dfrac{1}{|\xi-\omega|}\right)\dfrac{\xi-\omega}{2\omega} + \\ +\text{sign}(\xi+\omega)\dfrac{\ln((\xi+\omega)^2+1)}{4\omega} - \text{sign}(\xi-\omega)\dfrac{\ln((\xi-\omega)^2+1)}{4\omega}\end{array}\right). \quad (13d)$$

Rather cumbersome expressions for conic part contributions $p_3^L(\xi,\omega)$ and $p_1^L(\xi,\omega)$ are listed in the end of Appendix D.

Note, that the first term in Eq. (13a) is independent of $h_m$ and thus represents the "ideal image" of the domain wall, $\beta(\xi,\omega)$, while the second term is related with PFM object transfer function halfwidth. For small values, $h_m \ll b$, and the second term is negligible near the wall center (i.e. at $|y| \ll b$). At the same time, the dependence of PFM profile on the intrinsic width $b$ is negligible far from wall.

From Eqs. (12,13) the vertical response is zero at the center of the wall (at $y = 0$) and saturates far from the wall (at $|y| \to \infty$), while the lateral PFM response is maximal at the center of the wall and tends to zero far from wall. Note that at $b \to 0$, Eq. (12-13) derived for oblique-like flat domain wall reduce to Eqs.(9)-(10) for infinitely sharp domain wall as anticipated.

Vertical PFM response near the oblique domain wall in PbTiO$_3$ as a function of the distance from the wall is shown in Fig. 5.

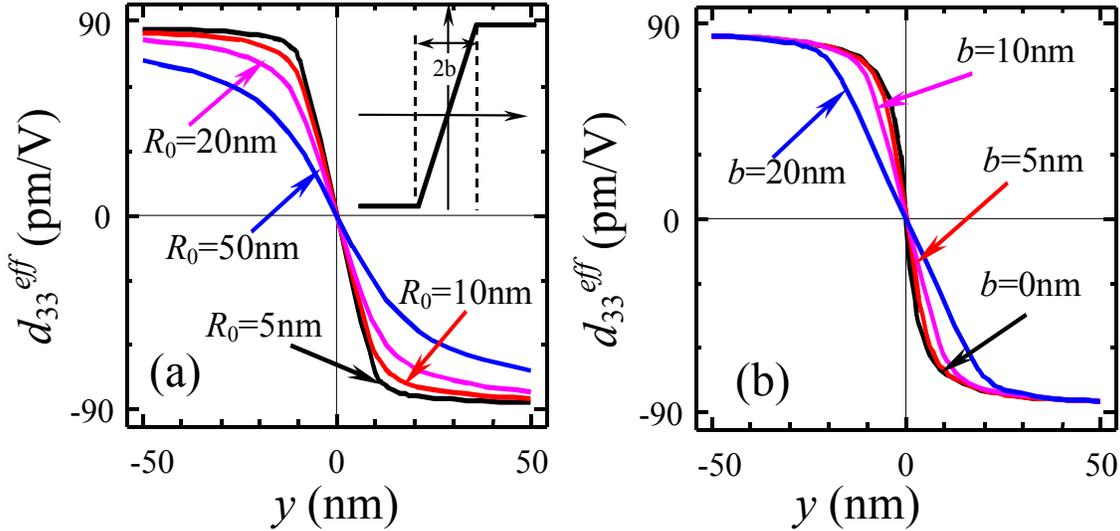

**FIG. 5.** (Color online) Vertical PFM response near the isolated domain wall in PbTiO$_3$ as a function of the distance from the wall (a) with intrinsic width $b = 10$ nm and different $R_0$ values (labels near the curves); (b) with $R_0 = 10$ nm and different intrinsic width $b$ values (labels near the curves). Piecewise smooth distribution like "oblique step" is shown schematically on inset. Other parameters are the same as in Fig. 3.

Numerical calculations prove that the relative contribution of conic part to the effective piezoresponse $d_{33}^{eff}(y)$ is negligible (about 1-5%) in the case $L/R_0 \ll \exp((\varepsilon_e + \kappa)/2\varepsilon_e)$, possible for typical ferroelectrics in air (since $\kappa \gg \varepsilon_e$), while the disk contribution strongly dominates (similarly to the case of infinitely sharp domain wall). However, the conic part affects the piezoresponse saturation rate. For perfect tip-surface electric contact the contribution of conical part is negligible in the vicinity of domain wall $|y| < 2R_0$, providing the validity of effective point charge approximation with high accuracy under the aforementioned conditions.

It is clear from Fig. 4, that intrinsic contribution to measured wall width becomes significant only for intrinsic width $b > R_0/2$. Moreover, both analytical results (12)-(13) and numerical simulations for different intrinsic domain wall profiles $\beta(x)$, lead to the same conclusion: the intrinsic domain width effect on the PFM response becomes negligible at $2b < h_m$ over all range of available material parameters. This rather general result along with the estimation of $h_m \sim 5-50$ nm (valid for standard conductive tips depending on their curvature and experiment geometry) and inequality $b < (2-5)$ nm (typical for bulk perovskites like PbTiO$_3$ or BaTiO$_3$,[29] LiTaO$_3$,[16] and Rochelle Salt[29]), corroborates the proposed tip calibration procedure elaborated in Ref. [30] for effective point charge approach and infinitely sharp domain wall approximation.

The opposite situation $b \geq 10$ nm may be realized in LiNbO$_3$ [18] and organic polar materials. Also one may suspect $h_m \ll 5$ nm for atomic or ultra-sharp tips with small curvature $R_0$. Under the condition $b > h$ both effective point charge-surface separation $h$ and intrinsic width $b$ should be taken into account for the accurate fitting of the domain wall PFM profiles as well as the aforementioned tip calibration procedure should be modified as discussed below.

## V. Discussion

Closed-form analytical expressions (11) and (12)-(13) for vertical and lateral PFM profiles of finite-width domain wall are derived. They take into account the conical and contact parts of the probe, and the material dielectric anisotropy $\gamma$. Similar to the case of an infinitely sharp domain wall, under the condition $L/R_0 \ll \exp((\varepsilon_e + \kappa)/2\varepsilon_e)$ the contribution of conical part is negligible in the vicinity of domain wall, providing the validity for effective point charge approximation with high accuracy.

In the effective point charge approximation of the tip,[31] electric field and dielectric anisotropy $\gamma \approx 1$, vertical piezoresponse at a distance $y$ from the oblique-like domain wall $\beta(y,b) = (|y+b|-|y-b|)/2b$ located at $y=0$ acquires the simplest form:

$$d_{33}^{eff}(y,b,h) = \begin{pmatrix} -\left(\left(\frac{1}{4}+\nu\right)d_{31} + \frac{3}{4}d_{33} + \frac{d_{15}}{4}\right)\frac{|y+b|-|y-b|}{2b} - \\ -\frac{d_{15}}{4}\frac{3h}{8b}\ln\left(\frac{4|y-b|+3h}{4|y+b|+3h}\right) - \\ -\left(\left(\frac{1}{4}+\nu\right)d_{31} + \frac{3}{4}d_{33}\right)\frac{h}{8b}\ln\left(\frac{4|y-b|+h}{4|y+b|+h}\right) \end{pmatrix} \quad (14)$$

Here the first term is the ideal image $\beta(y,b)$ of domain wall intrinsic profile. Eq. (14) can be used for two-parametric domain wall profile fitting and tip calibration, where the effective point charge-surface separation $h$ and intrinsic width $b$ are fitting parameters.

When the intrinsic width $b$ is several times smaller or greater than effective point charge-surface separation $h$, Eqs. (14) can be further simplified as:

$$d_{33}^{eff}(y,b,h) \approx -\beta(y,b)\left(\frac{|y|+b}{|y|+b+h/4}\left(\left(\frac{1}{4}+\nu\right)d_{31} + \frac{3}{4}d_{33}\right) + \frac{d_{15}}{4}\frac{|y|+b}{|y|+b+3h/4}\right) \quad (15)$$

Numerical calculations prove that Eq. (15) is approximately valid for an arbitrarily smooth domain wall profile $\beta(y,b)$, with characteristic intrinsic width $b$ (e.g. for $\tanh(y/b)$ or $(1-\exp(-|y|/b))\text{sign}(y)$). Thus, for a given $\beta(y,b)$ we could fit the measured domain wall profile $d_{33}^{eff}(y)$ and extract $h$ and $b$ values. Essential progress of closed-form expressions (11)-(13) and especially approximation (15) in comparison with more rigorous numerical fitting of experimental data demonstrated earlier,[18] consists in the possibility of simple and unique analytical interpretation of available experimental data. The disadvantage of Eq. (15) is the insufficient accuracy for accurate quantitative domain wall profile reconstruction, while qualitative analyses can be easily performed.

On the other hand, the first term in Eq.(15) is the ideal image $\beta(y,b)$ of domain wall intrinsic profile. For $b \to 0$ the second bracket is exactly the absolute value of the PFM response of an infinitely sharp domain wall [32] and the function $\beta(y,b) \to \text{sign}(y)$ as anticipated. Actually, Eq. (15) is the product of intrinsic (structural) factor and pseudo-extrinsic factor related with OTF features and intrinsic width $b$ superposition. Note, that we could not neglect $b$ value in the immediate vicinity of domain wall ($|y| \ll b$) even in the case $b \ll h$. Thus, the extrinsic factor appears naturally broadened with intrinsic width $b$, proving that intrinsic and extrinsic factors cannot be easily separated.

Within effective point charge approximation, both intrinsic width $b$ and effective tip parameter $h$ can be extracted from the measured piezoresponse profile slope in the immediate vicinity of domain wall (i.e. at $|y| \ll b$), and determining the piezoresponse saturation rate far from the wall (i.e. at $|y| \gg b+h$), since expansions exist

$$d_{33}^{eff}(y,b,h) \approx \begin{cases} -y\left(\dfrac{d_{15}}{4}\dfrac{1}{b+3h/4} + \left(\left(\dfrac{1}{4}+\nu\right)d_{31} + \dfrac{3}{4}d_{33}\right)\dfrac{1}{b+h/4}\right), & |y| \ll b \\ -\left(\left(\dfrac{1}{4}+\nu\right)d_{31} + \dfrac{3}{4}d_{33}\right)\left(1-\dfrac{h}{4y}\right) - \dfrac{d_{15}}{4}\left(1-\dfrac{3h}{4y}\right), & y \gg b+h \end{cases} \quad (16a,b)$$

Eq. (16) is valid with quite satisfactory for an arbitrarily smooth domain wall profile $\beta(y,b)$, with characteristic intrinsic width $b$. So, for material with known piezoelectric coefficients $d_{ij}$ the procedure for $h$ and $b$ determination is following:

1. At the first step the effective tip parameter $h$ is determined from the piezoresponse fitting far from the wall by Eq.(16b), since here $d_{33}^{eff}(y) \approx d_0 - d_1 h/y$.

2. At the second step the tangential slope $\alpha$ near the wall can be determined using least squire method and Eq.(16a), since here $d_{33}^{eff}(y) \approx \alpha y$. Then intrinsic width b can be obtained from the slope $\alpha$ as solution of quadratic equation.

To summarize, obtained analytical expressions provide insight into the mechanisms of domain structure PFM image formation. Namely, PFM profile of a realistic domain wall is the complex convolution of its intrinsic profile and extrinsic factors related with the non-locality of the PFM resolution function; it could not be reduced to their product even in the simplest cases. This result distinguishes PFM imaging from the "far-field" methods, where a typical diffraction pattern represents the product of structural and scattering factors.

## Appendix A. Domain wall profile

Using the basic relations Eqs. (1), (2) of the decoupled theory and expression (3) for the response near ideal isolated wall one can write the response near the sharp peak (Dirac-delta function distribution) in the form:

$$u_i^{delta}(y-y_0) = \int_{-\infty}^{\infty} d\xi_1 \int_{-\infty}^{\infty} d\xi_2 \, W_{ijkl}(-\xi_1,-\xi_2) d_{lkj} \delta(y-y_0-\xi_1) \equiv \frac{1}{2} \frac{\partial u_i^{step}(y-y_0)}{\partial y} \quad (A.1)$$

For the arbitrary one-dimensional distribution of $d_{lkj}(y_1-\xi_1) = d_{lkj} \cdot \beta(y_1-\xi_1)$ the response is

$$u_i(y_1) = \int_{-\infty}^{\infty} d\xi_1 \int_{-\infty}^{\infty} d\xi_2 \, W_{ijkl}(-\xi_1,-\xi_2) d_{lkj} \cdot \beta(y_1-\xi_1). \quad (A.2)$$

Using the Delta function definition $\beta(y_1-\xi_1) = \int_{-\infty}^{\infty} \delta(x-y_1+\xi_1)\beta(x)dx$, Eq.(A.2) can be rewritten in the form $u_i(y_1) = \int_{-\infty}^{\infty} dx \int_{-\infty}^{\infty} d\xi_1 \int_{-\infty}^{\infty} d\xi_2 \, W_{ijkl}(-\xi_1,-\xi_2) d_{lkj} \delta(x-y_1+\xi_2)\beta(x)$, so:

$$u_i(y_1) = \int_{-\infty}^{\infty} dx \, u_i^{delta}(y_1-x)\beta(x) \equiv \int_{-\infty}^{\infty} dx \frac{1}{2} \frac{\partial u_i^{step}(y_1-x)}{\partial y_1} \beta(x). \quad (A.3)$$

For the single domain wall $u_i^{step}(y_1 \to \pm\infty) \to \pm u_i^{inf}$ and $\beta(x \to \pm\infty) \to \pm 1$, using the integration over parts, one can rewrite integrals (A.3) as follows:

$$\int_{-\infty}^{\infty} dx \frac{1}{2} \frac{\partial u_i^{step}(y_1-x)}{\partial y_1} \beta(x) = -\frac{1}{2} \beta(x) u_i^{step}(y_1-x) \Big|_{-\infty}^{\infty} + \frac{1}{2} \int_{-\infty}^{\infty} u_i^{step}(y_1-x) \frac{\partial \beta(x)}{\partial x} dx$$

$$= \frac{1}{2} \int_{-\infty}^{\infty} u_i^{step}(y_1-x) \frac{\partial \beta(x)}{\partial x} dx \quad (A.4)$$

## Appendix B. Disk part

Metallic disk of radius $R_0$ biased with the potential $\varphi_0$ and located on the boundary between isotropic and anisotropic dielectric with permittivity values $\varepsilon$ and $\varepsilon_{11}, \varepsilon_{33}$ respectively create electric field with the following potential:

$$\varphi_D(z,\rho,R_0) = \varphi_0 \frac{2}{\pi} \arcsin\left( \frac{2R_0}{\sqrt{(R_0+\rho)^2 + (z/\gamma)^2} + \sqrt{(R_0-\rho)^2 + (z/\gamma)^2}} \right) \quad (B.1a)$$

here anisotropy factor is $\gamma = 1$ for isotropic dielectric ($z < 0$) and $\gamma = \sqrt{\varepsilon_{33}/\varepsilon_{11}}$ for anisotropic one ($z > 0$). Note that for $\gamma = 1$ Eq. (1a) becomes the potential of biased disk in vacuum.[33]

Using the Fourier-Bessel integral transformation, one can rewrite potential (1a) as follows (see e.g. Ref. [34], sec. 6.75, eq. 6.752):

$$\varphi_D(z, \rho, R_0) = \varphi_0 \frac{2}{\pi} \int_0^\infty dk\, J_0(k\rho) \frac{\sin(k R_0)}{k} \exp\left(-k \frac{|z|}{\gamma}\right) \tag{B.1b}$$

The potential of the point charge $q$ near the boundary between isotropic and anisotropic dielectric for $z > 0$, $\varphi_q = 2q \Big/ \left(4\pi\varepsilon_0 (\varepsilon_e + \kappa)\sqrt{\rho^2 + (z/\gamma + h)^2}\right)$, is rewritten as:

$$\varphi_Q(z, \rho, h) = \frac{q}{4\pi\varepsilon_0} \frac{2}{\varepsilon_e + \kappa} \Psi(z, \rho, h), \quad \Psi(z, \rho, h) = \int_0^\infty dk\, J_0(k\rho) \exp\left(-kh - k\frac{z}{\gamma}\right). \tag{B.2}$$

From Eqs. (B.1b), (B.2), the following relationship can be established:

$$\frac{\partial \varphi_D(z, \rho, R_0)}{\partial R_0} = \varphi_0 \frac{2}{\pi} \frac{\Psi(z, \rho, h + i R_0) + \Psi(z, \rho, h - i R_0)}{2}\bigg|_{h \to 0} \tag{B.3}$$

So, the point charge potential is related to the disk potential (B.1b) through the reversible *linear* transformation on space variables $z, \rho$:

$$\varphi_D(z, \rho, R_0) = \varphi_0 \frac{2}{\pi} \int_0^{R_0} d\tilde{a}\, \frac{\Psi(z, \rho, h + i\tilde{a}) + \Psi(z, \rho, h - i\tilde{a})}{2}\bigg|_{h \to 0} \tag{B.4}$$

Here the limits of integration are chosen so that to get the point charge potential (B.2) in the limit $R_0 \to 0$ (and $h \neq 0$). One should also take into account "capacitance" of the disk:

$$\varphi_0 \frac{2}{\pi} R_0 \to \frac{q}{4\pi\varepsilon_0} \frac{2}{(\varepsilon_e + \kappa)}.$$

The PFM response, for a point charge model is:[20]

$$u_i^q(\mathbf{y}, h) = \frac{q}{4\pi\varepsilon_0} \frac{2}{(\varepsilon_e + \kappa)} U_i^Q(\mathbf{y}, d)$$

$$U_i^q(\mathbf{y}, h) = \int_{-\infty}^\infty d_{mnk}(\mathbf{y} - \boldsymbol{\xi}) \left( \int_{z=0}^\infty c_{jlmn} \frac{\partial \Psi\left(\xi_3, \sqrt{\xi_1^2 + \xi_2^2}, h\right)}{\partial \xi_k} \frac{\partial}{\partial \xi_l} G_{ij}(\xi_1, \xi_2, \xi_3) d\xi_3 \right) d\xi_1 d\xi_2 \tag{B.5}$$

Hence, $U_i^q$ is the linear operator on $\Psi$ acting only on its coordinate part. Thus we can use transformation (B.4) in order to obtain closed form expression for PFM response on the electric field of the biased metallic disk:

$$u_i^D(\mathbf{0},\mathbf{y},R_0) = \varphi_0 \frac{2}{\pi} \int_0^{R_0} d\tilde{a} \left. \frac{U_i^q(\mathbf{0},\mathbf{y},h+i\tilde{a}) + U_i^q(\mathbf{0},\mathbf{y},h-i\tilde{a})}{2} \right|_{h \to 0} \quad (B.6)$$

In principle Eq. (B.6) allows obtaining integral representation for PFM response of different structures, namely three –fold (two-fold) integrals for PFM response near flat domain wall, transfer function, etc.

**Appendix C. Conical part**

Electric field of the conical part of the tip can be approximated by the field created by line segment charged with density $\lambda$ (see e.g. Ref.[24]). Thus one can write the following representation for the potential of the line segment:

$$\varphi_L(z,\rho,\Delta L) = \begin{cases} \dfrac{1}{4\pi\varepsilon_0 \varepsilon_e} \displaystyle\int_0^L \left( \dfrac{\lambda}{\sqrt{\rho^2 + (z+\Delta L + \tilde{z})^2}} + \dfrac{\varepsilon_e - \kappa}{\varepsilon_e + \kappa} \dfrac{\lambda}{\sqrt{\rho^2 + (-z+\Delta L + \tilde{z})^2}} \right) d\tilde{z}, & z < 0; \\[2ex] \dfrac{1}{4\pi\varepsilon_0} \dfrac{2}{\varepsilon_e + \kappa} \displaystyle\int_0^L \dfrac{\lambda}{\sqrt{\rho^2 + (z/\gamma + \Delta L + \tilde{z})^2}} d\tilde{z}, & z > 0. \end{cases}$$

(C.1)

Here $L$ is the tip length, line charge density $\lambda$ is proportional the applied voltage $\varphi_0$ and depends on half-angle of the cone $\theta$, $\Delta L$ is a distance of the lower end of the line to the sample surface.

For the limiting case $L \to \infty$ and $\kappa \gg \varepsilon_e$ (metallic sample) one can obtain that $\lambda = 4\pi\varepsilon_0 \varepsilon_e \varphi_0 / \ln((1+\cos\theta)/(1-\cos\theta))$ and $\Delta L = l/\cos\theta$, where $l$ is the separation between the tip apex and surface.[24] However, in the case of dielectric samples one can hardly use these expressions, since the isopotential surfaces deviate greatly from cone for the case $\kappa \sim \varepsilon_e$. Moreover, in the combined model ("disk plus line") field in the vicinity of the surface is determined mainly by the disk part and $\Delta L$ should be considered as some fitting parameter.

Also one should correct the potential of the disk to the potential on the sample induced by the line:

$$\varphi_L(0, \Delta L) = \frac{1}{4\pi\varepsilon_0} \frac{2\lambda}{\varepsilon_e + \kappa} \ln\left(1 + \frac{L}{\Delta L}\right) \tag{C.2}$$

Then we can start calculation of PFM response on the line field by set of the transformations: firstly one should replace distance with $\Delta L \to \Delta L + \tilde{z}$, charge with $Q \to \lambda \Delta L \tilde{z}$ and secondly integrate on $\tilde{z}$ as $\int_0^L d\tilde{z}$. In this way, using the definition for the displacement (7) and approximation (8) we obtained displacement components as follows

$$u_i^L(y, \Delta L, L) = \frac{\lambda}{2\pi\varepsilon_0(\varepsilon_e + \kappa)} g_{ijk}^L(y, \Delta L, L) d_{kj} \tag{C.3}$$

with

$$g_{3jk}^L(y, \Delta L, L) = \text{sign}(y) \sum_V g_{3jk}^V(\gamma, \nu) \ln\left(\left(1 + \frac{L}{\Delta L}\right) \frac{|y| + C_{3jk}^V(\gamma) \Delta L}{|y| + C_{3jk}^V(\gamma)(\Delta L + L)}\right), \tag{C.4a}$$

$$g_{1jk}^L(y, \Delta L, L) = \sum_L g_{1jk}^L(\gamma, \nu) \ln\left(\frac{|y| + C_{1jk}^L(\gamma)(\Delta L + L)}{|y| + C_{1jk}^L(\gamma) \Delta L}\right). \tag{C.4b}$$

**Appendix D.**

Constants $C_{ijk}$ and $f_{ijk}$ depend on dielectric anisotropy factor $\gamma$ only:

$$f_{351}(\gamma) = -\frac{\gamma^2}{(1+\gamma)^2}, \quad f_{333}(\gamma) = -\frac{1+2\gamma}{(1+\gamma)^2}, \quad f_{313}(\gamma) = -\frac{2}{1+\gamma}. \tag{D.1a}$$

$$C_{313} = \frac{1+\gamma}{8\gamma^2} {}_2F_1\left(\frac{3}{2}, \frac{3}{2}; 3; 1 - \frac{1}{\gamma^2}\right), \quad C_{333} = \frac{3(1+\gamma)^2 \, {}_2F_1\left(\frac{3}{2}, \frac{5}{2}; 4; 1 - \frac{1}{\gamma^2}\right)}{16\gamma^2(1+2\gamma)}, \tag{D.1b}$$

$$C_{351} = \frac{3(1+\gamma)^2 \, {}_2F_1\left(\frac{3}{2}, \frac{3}{2}; 4; 1 - \frac{1}{\gamma^2}\right)}{16\gamma^2} \tag{D.1c}$$

$$f_{133} = \frac{3}{8\gamma} {}_2F_1\left(\frac{1}{2}, \frac{3}{2}; 3; 1 - \frac{1}{\gamma^2}\right), \quad f_{113} = \frac{1}{\gamma} {}_2F_1\left(\frac{1}{2}, \frac{1}{2}; 2; 1 - \frac{1}{\gamma^2}\right), \quad f_{151} = \frac{2}{\pi} - f_{133} \tag{D.1d}$$

$$C_{113} = \frac{1}{(1+\gamma)^2 f_{113}}, \quad C_{133} = \frac{\gamma}{(1+\gamma)^3 f_{133}}, \quad C_{151} = \frac{(3+\gamma)\gamma^2}{2(1+\gamma)^3 f_{151}}. \tag{D.1e}$$

Where $_2F_1(a,b;c;d)$ is the hypergeometric function. In particular $C_{313}(1) = \frac{1}{4}$, $C_{333}(1) = \frac{1}{4}$, $C_{351}(1) = \frac{3}{4}$, $f_{351}(1) = -\frac{1}{4}$, $f_{333}(1) = -\frac{3}{4}$, $f_{313}(1) = -1$ and so for $s=q$

$$d_{33}^{eff}(y,b,h) = -\left(\left(\frac{1}{4}+\nu\right)d_{31} + \frac{3}{4}d_{33}\right)p_3^q\left(\frac{4y}{h},\frac{4b}{h}\right) - \frac{d_{15}}{4}p_3^q\left(\frac{4y}{3h},\frac{4b}{3h}\right) \tag{D.2}$$

Expressions for $p_3^L(\xi,\omega)$ and $p_1^L(\xi,\omega)$ are:

$$p_3^L(\xi,\omega) \approx \left( \begin{array}{l} \dfrac{|\xi+\omega|-|\xi-\omega|}{2\omega} + (|\xi+\omega|+1)\dfrac{\ln(|\xi+\omega|+1)}{2\omega\ln\left(1+\dfrac{L}{\Delta L}\right)} - (|\xi-\omega|+1)\dfrac{\ln(|\xi-\omega|+1)}{2\omega\ln\left(1+\dfrac{L}{\Delta L}\right)} + \\ + \left(|\xi+\omega|+1+\dfrac{L}{\Delta L}\right)\dfrac{\ln\left(|\xi+\omega|+1+\dfrac{L}{\Delta L}\right)}{2\omega\ln\left(1+\dfrac{L}{\Delta L}\right)} - \left(|\xi-\omega|+1+\dfrac{L}{\Delta L}\right)\dfrac{\ln\left(|\xi-\omega|+1+\dfrac{L}{\Delta L}\right)}{2\omega\ln\left(1+\dfrac{L}{\Delta L}\right)} \end{array} \right)$$
(D.3a)

$$p_1^L(\xi,\omega) \approx \left( \begin{array}{l} \left(|\xi+\omega|+1+\dfrac{L}{\Delta L}\right)\ln\left(|\xi+\omega|+1+\dfrac{L}{\Delta L}\right) - \left(1+\dfrac{L}{\Delta L}\right)\ln\left(1+\dfrac{L}{\Delta L}\right) - \\ -(|\xi+\omega|+1)\ln(|\xi+\omega|+1) \end{array} \right)\dfrac{\text{sign}(\xi+\omega)}{2\omega\ln\left(1+\dfrac{L}{\Delta L}\right)} -$$

$$-\left( \begin{array}{l} \left(|\xi-\omega|+1+\dfrac{L}{\Delta L}\right)\ln\left(|\xi-\omega|+1+\dfrac{L}{\Delta L}\right) - \left(1+\dfrac{L}{\Delta L}\right)\ln\left(1+\dfrac{L}{\Delta L}\right) - \\ -(|\xi-\omega|+1)\ln(|\xi-\omega|+1) \end{array} \right)\dfrac{\text{sign}(\xi-\omega)}{2\omega\ln\left(1+\dfrac{L}{\Delta L}\right)}$$

(D.3b)

[25] M. Abplanalp. *Piezoresponse Scanning Force Microscopy of Ferroelectric Domains*. Ph.D. thesis. Swiss Federal Institute of Technology, Zurich (2001).

[26] G.M. Sacha, E. Sahagún, and J.J. Sáenz. *J. Appl. Phys.* **101**, 024310 (2007).

[27] S.V.Kalinin, E.Karapetian, and M. Kachanov. *Phys. Rev. B* **70**, 184101 (2004).

[28] A.N.Morozovska, E.A.Eliseev, and S.V. Kalinin *J. Appl. Phys.* **102**, 074105 (2007).

[29] G.Catalan, J.F. Scott, A. Schilling, and J. M. Gregg, *J. Phys.: Condens. Matter*, **19**, 022201, (2007).

[30] S.V. Kalinin, S. Jesse, B.J. Rodriguez, E.A. Eliseev, V. Gopalan, and A.N. Morozovska. *App. Phys. Lett.* **90**, 212905 (2007).

[31] Starting from the tip model of point charge series, numerical fitting of the domain wall profiles in $LiNbO_3$ and $Pb(Zr,Ti)O_3$ converged to the single charge both in air and liquid ambient.[30]

[32] S.V. Kalinin, B.J. Rodriguez, S. Jesse, Y.H. Chu, T. Zhao, R. Ramesh, E.A. Eliseev, and A.N. Morozovska, *Proceedings of National Academy of Science of USA* **104**, 20204 (2007).

[33] L.D. Landau, and E.M. Lifshitz, Electrodynamics of Continuous Media, (Butterworth Heinemann, Oxford, 1980).

[34] I.S. Gradshteyn and I.M. Ryzhik, Table of Integrals, Series, and Products, 5th ed., edited by A. Jeffrey (Academic, New York, 1994).
25